\documentclass{revtex4}
\usepackage{verbatim}
\usepackage[pdftex]{graphicx}
\usepackage{microtype}
\usepackage{hyperref}
\usepackage{amsmath}
\usepackage{amssymb}

\newcommand{\mrm}{\mathrm}

\newcommand{\fvec}[1]{\widetilde{{#1}}}    

\begin{document}
\title{Optical frequency standards for gravitational wave detection using satellite Doppler velocimetry}
\author{Amar Vutha \\ \textit{Department of Physics and Astronomy, York University, 4700 Keele St, Toronto ON M3J 1P3, Canada} \\ avutha@yorku.ca}

\begin{abstract}
Gravitational waves imprint apparent Doppler shifts on the frequency of photons propagating between an emitter and detector of light. This forms the basis of a method to detect gravitational waves using Doppler velocimetry between pairs of satellites. Such detectors, operating in the milli-hertz gravitational frequency band, could lead to the direct detection of gravitational waves. The crucial component in such a detector is the frequency standard on board the emitting and receiving satellites. We point out that recent developments in atomic frequency standards have led to devices that are approaching the sensitivity required to detect gravitational waves from astrophysically interesting sources. The sensitivity of satellites equipped with optical frequency standards for Doppler velocimetry is examined, and a design for a robust, space-capable optical frequency standard is presented.
\end{abstract}

\maketitle

\section{Introduction}
The detection of gravitational waves (GWs) is one of the prime goals of gravitational physics today. Although there is tantalizing indirect evidence for the existence of gravitational waves \cite{Psaltis2008}, they have never been detected directly. At present, there are two working approaches to detecting GWs -- pulsar timing arrays (PTA) in the nanohertz GW frequency band \cite{Hobbs}, and terrestrial optical interferometers in the dekahertz band \cite{Pitkin2011}. These approaches, however, are insensitive to the astrophysically natural $\mu$Hz-mHz band of GW frequencies. PTAs are insensitive in this band due to the the long transit time between the pulsar and earth, and the need to average pulsar signals for a long duration to obtain adequate signal-to-noise ratios. Terrestrial interferometers, on the other hand, are affected by low-frequency noise sources which limit their operation below $\sim$ 1 Hz. Upcoming missions such as eLISA \cite{Amaro-Seoane2012}, building upon the LISA concept \cite{Pitkin2011,Crowder2005} of measuring the distance between distant inertial test masses using ultra-stable lasers, are expected to probe the mHz GW band. Other proposed approaches to detecting GWs in this frequency band include the use of very large atom interferometers \cite{Dimopoulos2008}.

A complementary approach to probing the $\mu$Hz GW band is to use satellite Doppler velocimetry (SDV). This approach derives from the pioneering analyses of Kaufmann \cite{Kaufmann1970}, and Estabrook and Wahlquist \cite{Estabrook1975}. We refer the reader to Smarr \emph{et al.}\ \cite{Smarr1983} for a description of one possible detection concept, and \cite{Armstrong2006} for a comprehensive recent review.  Kaufmann's original analysis \cite{Kaufmann1970} showed that a GW, interacting with light propagating from an emitter to a detector, imprints a small apparent Doppler shift on the light wave. This apparent Doppler shift is proportional to the difference in GW amplitudes at the emitter and detector (times an order unity angular factor). If this small frequency shift can be resolved, it offers an attractive approach to detecting GWs. The apparent Doppler shift imprinted by the GW is oscillatory at the GW frequency, and therefore it is necessary to resolve the small Doppler shift on a timescale less than the GW period so that the signal is not averaged away to zero. This places strong constraints on the performance of the frequency standards carried by the emitter and detector, requiring them to reach a high level of stability within a relatively short averaging time (1000s of seconds).

The purpose of this article is to point out that the demonstrated performance of present-day optical frequency standards makes the SDV approach competitive with complementary detection proposals for low-frequency gravitational waves. In the following, we first briefly review the physics of SDV, followed by an analysis of the performance achievable with optical frequency standards. We then present a design for a portable optical frequency standard that can be used as a component of SDV detectors of gravitational waves.

\section{Gravitational wave detection using satellite Doppler velocimetry}
Consider a gravitational wave with a wave-vector $k$ propagating along the $z$-axis in a region of space. The metric in this region is
\begin{equation}
ds^2 = dt^2 - (dz^2 + dx^2 + dy^2) - h_{+}[k(z-t)](dx^2 - dy^2) - 2h_{\times}[k(z-t)]dx dy,
\end{equation}
written in the transverse-traceless gauge.\footnote{We perform this analysis in units where $c=1$.} Here $h_+$ and $h_\times$ denote the amplitudes of the two polarizations of the GW. The cyclic coordinates of the action integral of a free particle (or photon) in this region are $x,y,z+t$, leading to the conservation of the momentum components $p_x, p_y,p_z+p_t$. In addition, a photon's four-momentum satisfies the null constraint $\fvec{p} \cdot \fvec{p} = 0$.\footnote{For simplicity of notation and analysis, we denote four-vectors as e.g.\ $d \fvec{x} = (dt; d\vec{x})$, and calculate all dot products using the metric.}

For a photon propagating with constant components $p_x = \alpha, p_y = \beta, p_z + p_t = \gamma$, the null condition $\fvec{p} \cdot \fvec{p} = 0$ implies that its four-momentum has the form
\begin{equation}
\fvec{p} = \left[ \frac{\gamma}{2} (1+ \delta) ;\alpha,\beta,\frac{\gamma}{2} (1- \delta) \right],
\end{equation}
where
\begin{equation}\begin{split}
\delta(z,t) & = \frac{\alpha^2 + \beta^2 + (\alpha^2 - \beta^2) h_+ + 2 \alpha \beta h_\times}{\gamma^2} \\
& = \tan^2 \theta \left( 1 + \cos 2\phi \, h_+ + \sin 2\phi \, h_\times \right)
\end{split}\end{equation}
with the constants $\alpha = p_0 \sin\theta \cos\phi, \beta = p_0 \sin\theta \sin\phi, \gamma = p_0 \cos\theta$ written in terms of spherical polar parameters $p_0,\theta,\phi$. 

\begin{figure}[h!]
\centering
\includegraphics[width=0.7\columnwidth]{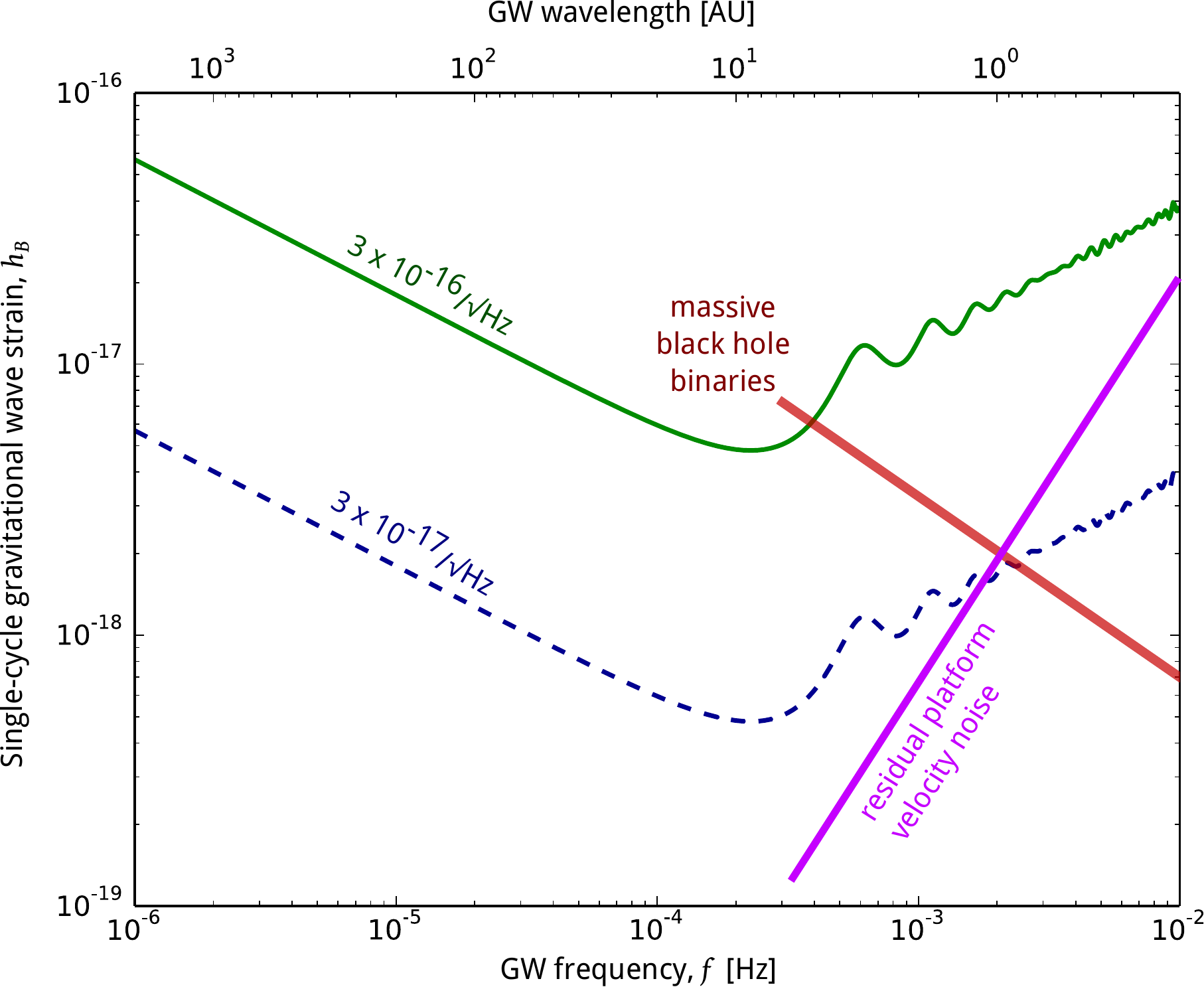}
\caption{\em \small Calculated response of a SDV GW detector, consisting of an emitter-receiver pair separated by 2 astronomical units (AU). The green solid line represents the single-cycle GW strain sensitivity (``burst sensitivity'') achievable with state-of-the-art atomic frequency standards \cite{Ludlow2014}, and the blue dashed line represents the performance achievable with a frequency standard that has ten-fold improved stability -- both lines are labeled by their SNR. Also shown is the limiting sensitivity due to residual velocity noise of the satellites, after corrections obtained from displacement sensing of an isolated inertial mass at the level of 0.5 $\mu$m/$\sqrt{\mathrm{Hz}}$. The red line is the theoretically predicted characteristic strain for GW radiation from one class of potential sources \cite{Moore2015}, showing that such sources are already detectable with state-of-the-art frequency standards.}
\label{fig:one}
\end{figure}

The photon's frequency measured by a free-falling emitter (or receiver) is $\nu = \frac{1}{2\pi} \fvec{p} \cdot (1;0,0,0) = \frac{\gamma}{4\pi} (1 + \delta)$. The ratio of the frequencies emitted by the emitter $e$, and detected at the receiver $r$, is therefore
\begin{equation}
\frac{\nu_\mrm{e}}{\nu_\mrm{r}}  = \frac{1 + \delta_\mrm{e}}{1 + \delta_\mrm{r}} \approx 1 + \sin^2\theta \left(\cos 2\phi \, \Delta h_+ + \sin 2\phi \, \Delta h_\times \right),
\end{equation} 
where $\Delta h_+ = h_+(\mrm{e}) - h_+(\mrm{r}), \Delta h_\times = h_\times(\mrm{e}) - h_\times(\mrm{r})$ are the differences between the GW strain amplitudes at the emitter and receiver. The frequency ratio $\nu_\mrm{e}/\nu_\mrm{r}$ is different from unity when $\Delta h_+$ or $\Delta h_\times$ is non-zero. The GW amplitude appears as an apparent Doppler shift at the receiver. It can be measured by precisely comparing the frequency of light received from the emitter to that of a stable frequency standard at the receiver. Therefore the fundamental limit to the performance of such a GW detector is the stability of the frequency standard carried by the emitter and receiver, evaluated on a timescale corresponding to the period of the GW. Figure \ref{fig:one} shows the single-cycle GW sensitivity of an emitter-receiver pair on board satellites in solar orbit (here onwards a ``GW detector''), for various values of the stability of the frequency standard. The performance of a SDV detector has a slower low-frequency roll-off compared to laser phase-measurement-based detection methods, and therefore offers a complementary method for low-frequency GW physics. (For the parameters chosen in Figure \ref{fig:one}, both the SDV response curves are more sensitive than the proposed eLISA detector \cite{Amaro-Seoane2012} in the 1-100 $\mu$Hz GW band.)

As was pointed out by Kaufmann \cite{Kaufmann1970}, and Estabrook and Wahlquist \cite{Estabrook1975}, note that a detectable frequency shift is obtained in spite of the fact that there is no GW perturbation on the local clock rates at the emitter and receiver (the $t$-component of the metric is unaffected by the GW). The imprinting of the GW's amplitude (which appears in the spatial part of the metric) onto the frequency of the photon is a consequence of the propagation of the massless photon through the perturbed spacetime.

We note that the stability of the frequency standard and its high signal-to-noise ratio (SNR) are crucial, not its absolute accuracy. Further, well-characterized long-term drifts may also be tolerated, as the GW signal appears as an oscillatory apparent Doppler shift at a characteristic Fourier frequency within the sensitivity band of the GW detector. This opens up the possibility of using frequency standards that may not necessarily be the best for time-keeping applications, but instead trade absolute accuracy for portability, stability and SNR.

The satellites will experience significant gravitational redshifts from being at different locations in the solar system's gravitational potential (making their absolute accuracy moot anyway), which will lead to slowly varying frequency shifts. In addition, effects such as perturbations from planets will lead to time-varying frequency shifts due to their effect on the satellite trajectories. However, these effects will typically appear on slower timescales ($\sim$months for orbital perturbations) compared to the $\sim$hour oscillation period of GW in the $\mu$Hz-mHz band. These spurious effects can therefore be clearly distinguished from GW signals.

The primary source of noise in the Doppler signal arises from uncontrolled motion of the satellite platform (with an approximately white velocity spectral density of $\sim 10^{-6}$ m/s/$\sqrt{\mrm{Hz}}$ at $\mu$Hz-mHz frequencies), due to the effects of solar wind, radiation pressure, and the spacecraft's mechanical imperfections \cite{Armstrong2006}. We suggest a method to compensate for this effect, using the fact that the resultant displacement noise is enhanced by a factor $\propto 1/f$: the displacement of an isolated free-falling test mass on board the satellite can be monitored, and the displacement record communicated between the satellites. This record is processed to correct the measured Doppler signals against the fluctuations in the relative velocity of the satellite platforms. By monitoring the displacement of the on-board inertial mass e.g.\ capacitive or optical interferometric sensing, the velocity noise can be compensated. With a displacement measurement sensitivity of 0.5 $\mu$m/$\sqrt{\mathrm{Hz}}$ (2 orders of magnitude less critical than the demonstrated performance of the drag-free control system in the LISA Pathfinder, in a similar frequency band \cite{Armano2009}), the velocity noise can be compensated to $< 10^{-9}$ m/s/$\sqrt{\mathrm{Hz}}$ at 100 $\mu$Hz. At this level, the residual platform velocity noise is no longer a significant concern for SDV detection, as shown in Figure \ref{fig:one}.

\begin{figure}[h!]
\centering
\includegraphics[width=0.5\columnwidth]{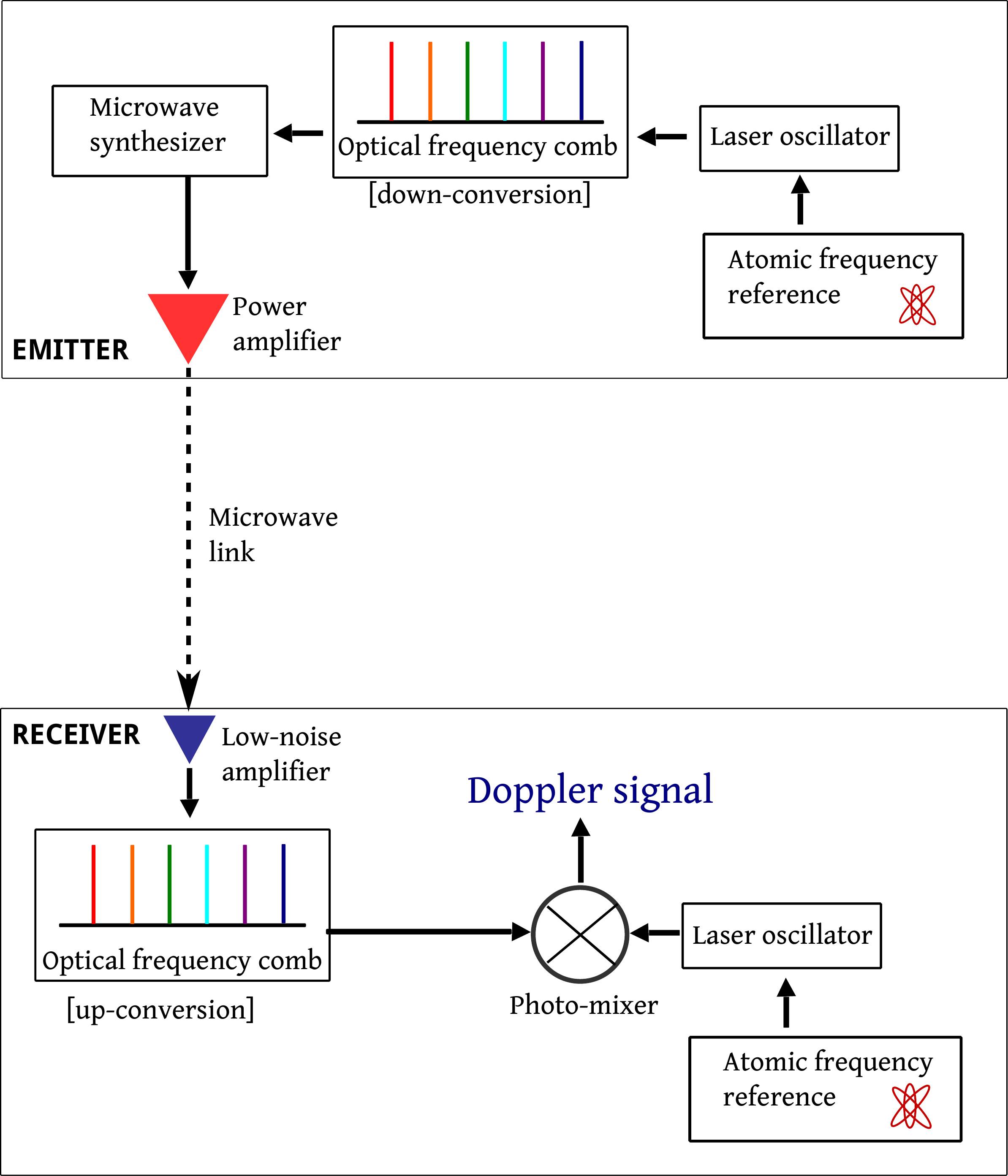}
\caption{\em \small {An example block diagram of a satellite Doppler velocimetry GW detector, with a simple one-way single-frequency link. In the emitter, a microwave frequency is obtained from the optical frequency standard, and transmitted over the link to the detector. At the receiver, the received microwaves are amplified, multiplied up to the optical domain and compared against the receiver's optical frequency standard. Multi-path links and multi-satellite links can be used to enhance the GW detection sensitivity \cite{Smarr1983}, and multiple-frequency links can be used to suppress the effects of plasma dispersion \cite{Armstrong2006}.}
} 
\label{fig:three}
\end{figure}

\section{Optical frequency standards}

Buoyed by the invention of the optical frequency comb \cite{Maddaloni2009}, optical frequency standards have made enormous leaps in performance over the last decade, and are presently the best frequency standards available \cite{Margolis2014,Ludlow2014}. These devices generate a stable optical frequency by slaving a laser oscillator to a high-quality-factor atomic resonance. The current state-of-the-art is achieved by optical lattice clocks, where the reference atoms are cooled and trapped in standing waves of light \cite{Ludlow2014}. Optical frequency standards are tremendously attractive as components of a SDV GW detector. Figure \ref{fig:three} illustrates one possible method for performing frequency comparisons between the emitter and receiver in a GW detector equipped with optical frequency standards.

At the emitter, a frequency comb is locked to an optical frequency standard (laser oscillator stabilized to an atomic frequency reference). The microwave frequency, obtained by dividing down the optical frequency using the frequency comb \cite{McFerran2005}, is transmitted over the inter-satellite link to the detector. At the receiver, the received microwaves are amplified by a low-noise amplifier, multiplied up to the optical domain using a frequency comb, and the resulting optical frequency compared to the receiver's frequency standard. The microwave link can be operated with multiple frequencies to cancel the effect of plasma dispersion \cite{Iess1999,Armstrong2006}. Alternatively, it is foreseeable that a high signal-to-noise ratio optical link can be operated between satellites in the near future \cite{Tolker-Nielsen2002,Hemmati2006}, allowing direct comparisons between optical signals from the emitter and receiver. 

Most of the components involved in this scheme are compatible with operation on board a satellite, using currently available technology. This includes miniaturized optical frequency combs that have been developed in recent years \cite{Papp2014}. The notable exception is the frequency standard, as optical frequency standards capable of being used on satellites have yet to be developed. In the next section, we address the feasibility of building and operating such a frequency standard.

\subsection{Towards a satellite-compatible optical frequency standard}
Due to their size and complexity, the state-of-the-art architecture for optical frequency standards (optical lattice clocks) may not be the best option for satellite-borne frequency standards (although cf. the efforts to miniaturize optical lattice clocks for measurements on the International Space Station \cite{Chen2014a}). Other more portable designs, such as single-ion clocks \cite{Jau2012} or solid-state clocks \cite{Rellergert2010}, might be better suited to being operated on a satellite. One of the main sources of the complexity of state-of-the-art optical frequency standards arises due to the long cycle times involved in interrogating the atomic reference using the laser oscillator. A high-performance laser oscillator, often with a linewidth of $\lesssim$ 1 Hz, must be used to interrogate the atomic transition for many seconds and the atoms must be held in place for the duration of the interrogation. By relaxing the constraints on the atomic frequency reference and reducing the measurement cycle time, the engineering of the frequency standard can be simplified. In this section, we present an alternative design for a portable optical frequency standard that could be useful as a component in a SDV GW detector. Satellites carrying high-peformance frequency standards would also be useful in precisely measuring gravitational properties of the solar system (such as gravitational redshifts) and constraining models of modified gravity.

\begin{figure}[h!]
\centering
\includegraphics[width=0.5\columnwidth]{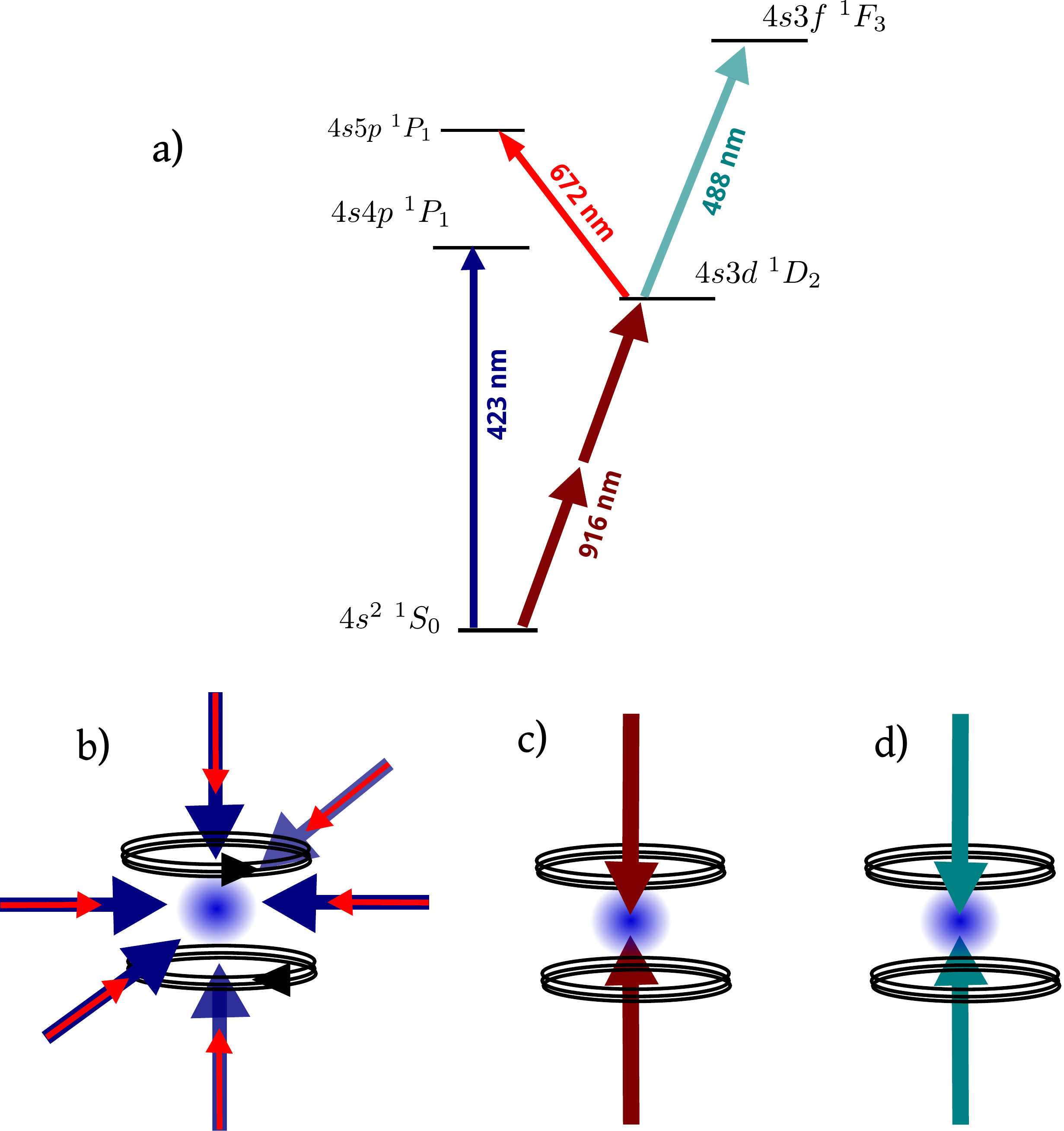}
\caption{\em \small The two-photon $^1S_0 - ^1D_2$ clock transition in neutral calcium (916 nm, 327 THz) is an attractive candidate for operating a simple optical frequency standard. The relevant transitions are shown in (a), and all the required laser wavelengths can be obtained from compact laser diodes. The measurement sequence involves the following steps: (b) atoms are cooled and trapped in a magneto-optical trap (MOT) operating on the 423 nm $^1S_0 - ^1P_1$ transition, (c) they are excited out of the $^1S_0$ state by two counter-propagating photons obtained from the 916 nm laser oscillator, and (d) the excited state fraction is detected using laser-induced fluorescence on the $^1D_2 - ^1F_3$ cycling transition at 488 nm. }
\label{fig:four}
\end{figure}

We have studied the design of an atomic frequency standard using the 916 nm $^1S_0 - {}^1D_2$ two-photon transition in neutral calcium atoms. This transition has a linewidth $\Gamma \simeq$80 Hz, which relaxes some of the constraints on the laser oscillator required to interrogate it, and leads to a faster cycle time. The relevant atomic states and transition wavelengths in calcium are shown in Figure \ref{fig:four}. The choice of this atomic reference also means that the frequency standard can be built using a single-stage magneto-optical trap (MOT), without incurring the complexity of the multiple stages of laser cooling and trapping required in optical lattice clocks. In contrast to schemes using single photon excitation, the thermal motion of the atoms in the MOT does not lead to significant frequency shifts in this design, due to the Doppler-free and recoil-free nature of degenerate two-photon excitation. The measurement scheme is compatible with low-power operation using robust and inexpensive diode lasers and a simple vacuum apparatus, and could lead to a portable and maintenance-free atomic frequency standard. We estimate that a frequency stability $\delta \nu/\nu \approx 10^{-16} / \sqrt{\tau}$ can be obtained with an integration time of $\tau$ seconds, with commensurate control over systematic frequency shifts. 

\section{Summary}
The high performance of modern optical frequency standards can be leveraged in order to detect gravitational waves using satellite Doppler velocimetry. The performance of such GW detectors is comparable to, and complements, proposed space-based interferometric detectors, but relies on a simpler approach based on demonstrated technologies. The feasibility of operating such detectors hinges on the availability of portable and maintenance-free optical frequency standards, and we have described one path to constructing such a device. The development of satellite Doppler velocimetry for GW detection could offer a unique window into the gravitational wave spectrum, and address important open questions in gravitational physics.

\section*{Acknowledgments}
I am grateful to Matt Johnson for very helpful discussions regarding gravitational wave sources. I also acknowledge valuable discussions with Luis Lehner and Asimina Arvanitaki during a visit to the Perimeter Institute. This work is supported by a Society in Science Branco Weiss Fellowship, administered by the ETH Zurich.

~\\ 
\emph{Note added}: Recently, somewhat similar proposals have been reported \cite{ripoff,ripoff2}.

\bibliography{gw_detection}

\end{document}